\documentclass{emulateapj}

\begin{document}

\title{Spin alignment of dark matter haloes in filaments and walls}

\author{ Miguel A.~Arag\'on-Calvo\altaffilmark{1}, Rien van~de~Weygaert\altaffilmark{1},
         Bernard J.~T.~Jones\altaffilmark{1} and J.M.~(Thijs)~van~der~Hulst\altaffilmark{1} }

\altaffiltext{1}{Kapteyn Astron. Inst., University of Groningen, PO
Box 800, 9700 AV Groningen, The Netherlands; miguel@astro.rug.nl}
 
\submitted{Submitted to ApJL}

\begin{abstract}
The MMF technique is used to segment the cosmic web as seen in a
cosmological N-body simulation into wall-like and filament-like
structures.  We find that the spins and shapes of dark matter haloes
are significantly correlated with each other and with the orientation
of their host structures.  The shape orientation is such that the halo
minor axes tend to lie perpendicular to the host structure, be it a
wall or filament.  The orientation of the halo spin vector is mass
dependent.  Low mass haloes in walls and filaments have a tendency to
have their spins oriented within the parent structure, while higher
mass haloes in filaments have spins that tend to lie perpendicular to
the parent structure.
\end{abstract}


%--------------------------------------------------------------------------
%     Intro
%--------------------------------------------------------------------------
\section{Introduction}
The origin of the angular momentum of galaxies and their associated dark
matter haloes remains one of the most poorly understood subjects in present
galaxy formation theories despite its relevance in determining properties
such as size and morphological type. According to the Tidal Torque Theory (TTT),
galaxies acquire their angular momentum as a consequence of the tidal shear
produced by the neighbouring primordial matter distribution \citep{Hoyle49,
Peebles69,Doroshkevich70,White84}. A natural consequence of the
TTT is a correlation between the angular momentum of haloes and their surrounding
matter distribution \citep{Efstathiou79,Barnes87,HeavPeac88,Lee01,Porciani02}.
\citet{Lee01} and \citet{Lee04} made specific predictions about this.
 
Orientation studies based on galaxy catalogues show anti-alignment \citep{Kashikawa92,
Navarro04,Trujillo06}. The situation in N-body models is less clear: some
dark matter N-body simulations seem not to detect any systemic halo spin alignment
\citep{Patiri06,Heymans06}, while others \citep{Lee01, Bailin05, Hatton01, Faltenbacher02}
present evidence that dark matter haloes are aligned with host structures.
More recently \cite{Altay06} found a strong shape alignment of dark haloes 
in filaments, while \cite{Trujillo06b} reported the discovery of a systematic 
alignment effect in an analysis of dark matter haloes taken from the very large ``Millennium'' N-body
simulation~\citep{millsimul05}. We suggest that such ambiguities may in large part be a consequence
of the methods used to identify the larger scale structures that host the haloes.
 
%--------------------------------------------------------------------------
%     Finding and classifying structure
%--------------------------------------------------------------------------
\section{Finding and classifying structure}
Large galaxy surveys such as the 2dF \citep{Colless01}
or more recently the Sloan Digital Sky Survey \citep{York00} have unambiguously revealed
an intricate network of galaxies: the cosmic web.  The cosmic web can be
described as a mixture of three basic morphologies: clusters which are predominantly spherical,
elongated associations of galaxies (filaments) and large planar structures (walls)
\citep{zeldovich70, shandzeld89}.
 
%-------------------
%----  FIGURE  -----
%-------------------
\begin{figure*}[!htp]
  \centering
  \includegraphics[width=0.95\textwidth,angle=0.0]{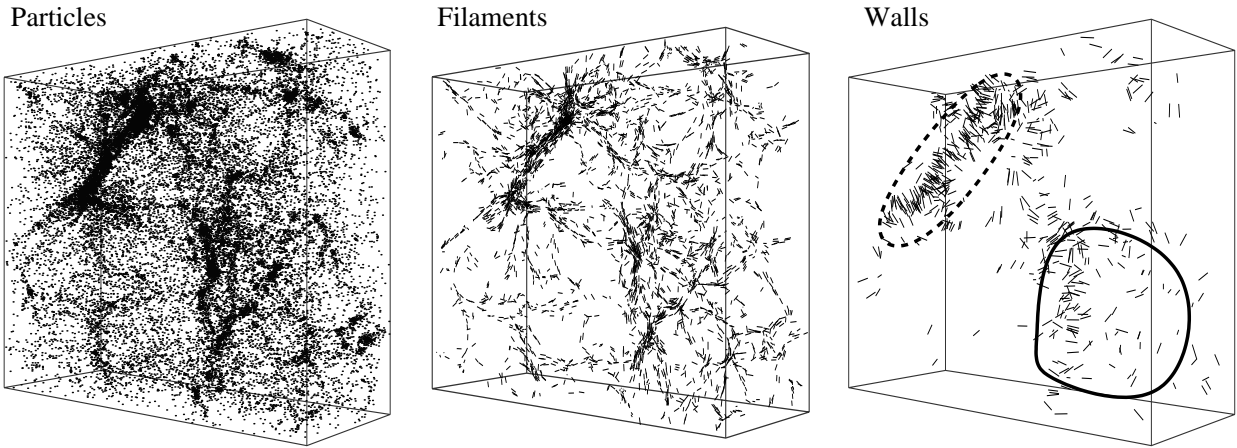}
    \caption{Left panel: Particles inside a sub-box of $37.5 \times 75 \times 100$ h$^{-1}$ Mpc.
      For reasons of clarity only a small fraction of the total number of particles is shown.
      Central panel: filaments delineated by a subsample of the particle distribution.
      At each particle location we have plotted the filament vector $\mathbf{e_F}$, indicating the
      direction locally parallel to the filament. Righthand panel: wall particles detected
      in the same sub-box: at each wall particle we plot the wall vector $\mathbf{e_W}$.
      Two walls can be clearly delineated: one seen edge-on (dashed outline) and one seen
      face-on (solid outline).}
  \label{fig:N_body_all}
\end{figure*}
 
\cite{bondweb96} emphasized that this weblike pattern is shaped by the large scale tidal force fields
whose source is the inhomogeneous matter distribution itself (see also \cite{weywhim05}).
Acccording to TTT the same tidal field also generates the angular momentum of collapsing halos.
Thus we would expect the shape and angular momentum of cosmic haloes to be correlated with
one another and with the cosmic web elements in which they are embedded.
 
Revealing such correlations requires the ability to unambigiously identify the structural features
of the cosmic web.  Several methods have in the past been used in an attempt to identify and
extract the morphological components of the observed galaxy distribution
\citep{Barrow85,Babul92,Luo95,Stoica05,Colberg05,Pimblet05} with
varying degrees of success. The results presented in this letter are based on a new method, 
the Multiscale Morphology Filter: ``MMF'' \citep{miguel06}.
MMF allows us to objectively segment the cosmic web into its three basic
morphological components by analysing the properties of the matter distribution hierarchically. 
With this morphological characterisation
we can isolate specific host environments (filaments and walls) for haloes
and test predictions from the TTT in a systematic way.  
 
The MMF method is a significant advance on other similar studies of N-body models.
MMF is a technique that locates and classifies various structures by exploiting localised 
properties of the inertia tensor of the matter distribution on a hierarchy of scales.
Since the inertia tensor is directly related to the dynamical forces that drive the tidal
torques, MMF is particularly suited for this investigation.
The significance of the effects reported here is strong, despite the relatively small 
size of the N-body simulation, because of the clear cut MMF environmental descriptor.

%--------------------------------------------------------------------------
%     N-body simulations and halo catalogue
%--------------------------------------------------------------------------
\section{N-body simulations and structure}
We ran a cosmological N-body simulation containing $512^3$  equal mass
dark matter particles inside a cubic box of $150$ h$^{-1}$ Mpc. using the public version 
of the parallel Tree-PM code Gadget2 \citep{Springel05}. We adopted the standard cosmological
model $\Omega_m=0.3$, $\Omega_\Lambda=0.7$, $h=0.7$ and $\sigma_8=1$.
The analysis presented here is based on the last snapshot at $z=0$. The mass
per particle is $2\times10^{9}$ h$^{-1}$ M$_{\odot}$  and the softening length
was set to 18 h$^{-1}$ kpc (comoving) until $z=2$ and 6 h$^{-1}$ kpc (physical)
after that time.
%--------------------------------------------------------------------------
%     MMF
%--------------------------------------------------------------------------
\subsection{Detecting Filaments and Walls}
The Multiscale Morphology Filter is based on the
second-order local variations of the density field as encoded in the Hessian
matrix ($\partial^2 \rho / \partial x_i \partial x_j$) for the smoothed density field. 
For a given set of smoothing scales we compute the eigenvalues of
the Hessian matrix at each position on the density field.
The density field is computed from the particle distribution by means of the DTFE
method~\citep{schaapwey2000, schaapwey06}: this is natural and self-adaptive,
and retains the intricate anisotropic and hierarchical features characteristic of the cosmic web.
{\it DTFE smoothing plays an essential role in delineating filaments and walls unambiguously}.
%-------------------
%----  FIGURE  -----
%-------------------
\begin{figure*}[!htfb]
  \centering
  \includegraphics[width=1.0\textwidth,angle=0.0]{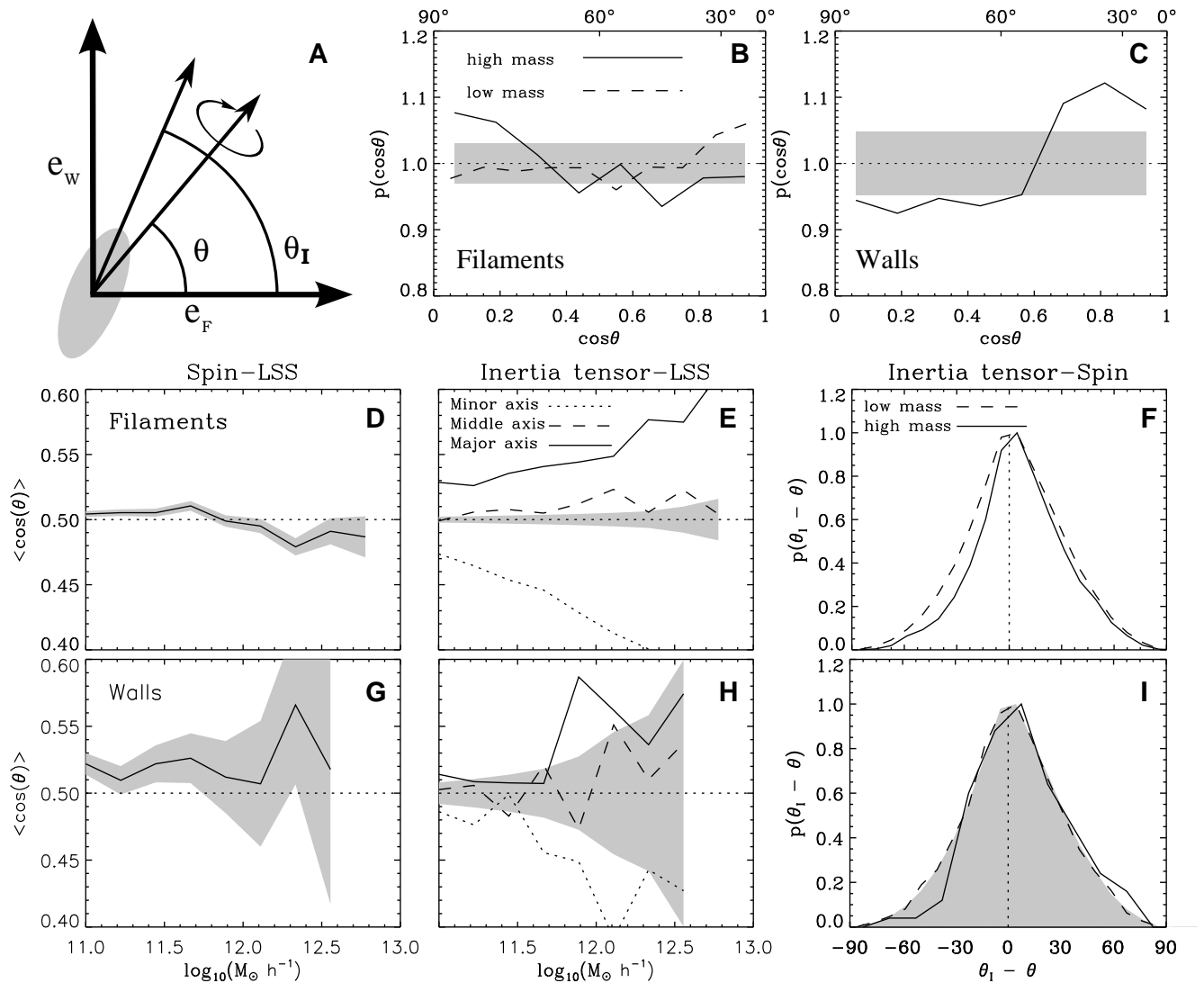}
    \caption{
{\bf A:} cartoon illustrating the parameters describing the relative orientations of halos,
filaments and walls .
{\bf B:} The probability distribution of $\cos\theta$: $\theta$ is the angle between the spin 
vector of a dark matter halos and its host filament. A distinction has been made between haloes with
masses smaller and larger than $10^{12}$ h$^{-1}$ M$_{\odot}$ (dashed vs. solid line).
The dotted line indicates a uniform distribution of halo orientations. The shaded area corresponds
to the standard deviation of 1000 random realisations with the same number of galaxies as the
halo sample. The same prescription is followed to estimate and indicate the standard deviation in the
other frames. 
{\bf C:} Similar to B, for walls: the pdf for the halo spin orientation $\theta$ wrt.
the embedding wall. 
{\bf D:} The angle $\cos\theta$ between the halo spin direction and
the host filament as a function of halo mass $M$. The solid line indicates the average
$\langle\cos\theta\rangle$ for the halo sample. The dotted line is the expected distribution for
a randomly oriented sample. 
{\bf G:} Similar to D, for walls. Note that the shaded deviation band is wider
as a result of the lower number of wall haloes. 
{\bf E:}  The alignment between halo shape and
host filament: the average orientation angle $\langle \cos\theta \rangle$ of major (solid line),
medium (dashed line) and minor (dotted line) halo axes relative to the filament. 
{\bf H:} Similar to E, for walls. 
{\bf F:} The pdf of the angle between halo minor axis and halo spin axis for haloes
in filaments. The figure distinguishes between low mass (dashed) and high mass haloes (solid). 
{\bf I:}
The pdf of the angle between halo minor axis and halo spin axis, for haloes in walls. Superimposed
(shaded area) is the same distribution for low mass haloes in filaments taken from F.}
  \label{fig:spin_alignment_ALL}
\end{figure*}
We use a set of morphology filters based on relations between the eigenvalues in order to get 
a measure of local spherical symmetry, filamentariness or planarity. The morphological 
segmentation is performed in order of increasing degrees of freedom in the
eigenvalues for each morphology (i.e. blobs $\rightarrow$ filaments $\rightarrow$ walls). 
The response from the morphology filters computed at all scales is
integrated into a single multiscale response which encodes the morphological
information present in the density field. At each stage of the filtering
we discard the particles that have previously been assigned to a structure and compute a new
density field. 
 
In a filament, the eigenvectors of the Hessian matrix
corresponding to the smallest eigenvalue ($\mathbf{e_{F}}$) indicate the
direction of filament.  In walls the largest eigenvalue, indicates the perpendicular 
to the wall ($\mathbf{e_{W}}$).
Eigenvectors are computed from a smoothed version  of the density field; this
avoids small-scale variations in the directions assigned to filaments and walls.
 
In Figure ~\ref{fig:N_body_all} we show a region of the simulation containing several filaments
and two large walls detected using the MMF. The box and its projection were chosen
in order to avoid confusion from projection of many structures on top of each other
and to show one wall face-on and one edge-on.
Filaments are clearly delineated like streaming lines joining large associations
of matter. Walls are more difficult to visualise so
we also plot their defining eigenvectors.
For this projection we can see that the eigenvectors
($\mathbf{e_W}$) defining the wall seen head-on (solid line) are pointing 
towards us, while the ones corresponding to the wall seen edge-on are perpendicular 
to the line-of-sight (dashed line).
 
%--------------------------------------------------------------------------
%        Halo Identification                                                                  
%--------------------------------------------------------------------------
\subsection{Halo Identification}
Haloes were identified with a somewhat different implementation
of the publicly available halo finder HOP \citep{Eisenstein98}.
First we identify haloes by running \verb|hop| with the standard parameters
and $\delta_{out}$=80, $\delta_{saddle}$=120 and $\delta_{peak}$=160 for
\verb|regrouping|. Each of these haloes is considered a parent candidate which may comprise
more than one single subhalo. Next for all particles we compute densities using a
Gaussian window
with dispersion of 35 h$^{-1}$ kpc, in order to produce a smoothed density field without
substructure smaller than this kernel. We run \verb|hop| again but only for particles
inside the parent candidates and this time we also provide the smoothed densities
previously computed as an input for \verb|hop|. The halo identification is performed without
running \verb|regroup|. By doing this \verb|hop| will assign all particles to their smoothed
local maximum, each group found in this stage is a candidate subhalo. For each of these
new subhaloes we find their center of mass iteratively and remove the unbound particles.
This ``FracHop'' method allows us to find bound subhaloes inside larger groups otherwise
identified as single virialised objects.
 
Finally, we keep haloes with more than 50 particles
and less than 5000, a mass range of (1-100)$\times10^{11}$ h$^{-1}$ M$_{\odot}$.
 
%--------------------------------------------------------------------------
%        Halo properties
%--------------------------------------------------------------------------
\subsection{Halo properties}
The angular momentum of a halo containing $N$ particles is then defined as:
\begin{equation}
\mathbf{J} = \sum_i^N m_i \; \mathbf{r_i} \times (\mathbf{v_i}-\mathbf{\bar{v}})
\end{equation}
\noindent where $m_i$ is the particle's mass, $\mathbf{r_i}$ is the
distance of each particle from the center of the halo, $\mathbf{v_i}$ is the
peculiar velocity of the particle and $\mathbf{\bar{v}}$ the mean
velocity of the halo with respect to the center of mass. We compute the angle
between the halo's spin and its assigned Filament or Wall,
\begin{equation}
\theta_W = 90-\phi_W, \quad \theta_F = \phi_F,
\end{equation}
where
\begin{equation}
\phi_{F,W} = \cos^{-1} \left( \frac{\mathbf{J} \cdot \mathbf{e_{F,W}} }
                     {\mid \mathbf{J}\mid \mid \mathbf{e_{F,W}} \mid} \right)\,,
\end{equation}
 
\noindent $\mathbf{e_{F,W}}$ being the vector defining the orientation of the
filaments (F) and walls (W). For each halo principal axes are computed by
diagonalising the inertia moment tensor
\begin{equation}
\mathbf{I}_{ij} = \sum_i^N\,m_i\,{x_i\,x_j},
\end{equation}
\noindent where the positions of the particles are with respect to the center of mass and
the sum is over all particles in the halo. 
The orientation of any of the eigenvectors of
$\mathbf{I}_{ij}$ relative to the host structure is described by angles analogous
to the angles $\phi$, $\theta_W$, and $\theta_F$ defined above.
 
%--------------------------------------------------------------------------
%        Results
%--------------------------------------------------------------------------
\section{Results}
Our main results are presented in Figure ~\ref{fig:spin_alignment_ALL} and may be summarised as follows:
\begin{itemize}
\item The orientations of halo spins are significantly
correlated with the large scale structure in which
they are situated. Haloes in walls have spin vectors that
tend to lie in the plane of the host wall (frame C).
The strength of the alignment of haloes in
walls is relatively small (the Lee-Pen correlation parameter
is $c=0.13 \pm 0.02$) but nevertheless statistically significant:
the K-S probability that the halo orientations in walls are
randomly distributed is less than
$8.16\times 10^{-5}$. We find similar correlations in
filaments, though of a more subtle character (frame B).
\item The orientation of halo spin vectors in filaments
is mass dependent: low mass halo spins tend to lie
parallel to their host filament while high mass halo
spins tend to lie perpendicular to their host filament (frames B \& D).
High mass haloes, $M>10^{12} h^{-1} M_{\odot}$, have a K-S
probability less than $5.50\times 10^{-4}$ of being randomly
orientated, low mass haloes less than $1.86\times 10^{-5}$ (frame B).
\item The principal axes of haloes in filaments are strongly
correlated with the direction of the filaments: the minor axis
tends to be directed perpendicular to the filament, the effect
being strongest for the larger masses (frame E)
\item The principal axes of haloes in walls are strongly
correlated with the perpendicular to walls: the minor axis
tends to be lie perpendicular to the wall while the other
axes tend to lie in the local plane of the wall (frame H).
\item While both the halo minor axis and the spin are
correlated with the megaparsec scale structure, the distribution of
the angle between these two is skewed and biased towards the
surrounding large scale structure. This slight skewness may bias
the estimate of the mean angle the spin makes with the host
structure (frames F and I).
\end{itemize}
%--------------------------------------------------------------------------
%        Concluding remarks
%--------------------------------------------------------------------------
\ \\
\section{Concluding remarks}
The significance of the alignments we have found, given the relatively small sample of haloes and
small simulation volume, emphasises the benefits of having a good definition of
environmental structures when studying halo properties and their relation with the cosmic web.
 
The results reported here are in accord with results reported by e.g. \cite{Trujillo06b}. 
We find that halo spin vectors lie within the host structure and halo minor axes point out 
of the host structure as previously suggested by \citep{Patiri06, Trujillo06b}. 
Th\'e major advance of our study is the objective and multiscale character of the 
structure identification by means of the MMF. Earlier studies at best resorted 
to heuristic means of delineating filaments or walls \citep{Pimblet05, Colberg05, 
Trujillo06b}. Some earlier studies have measured the distance of a halo from a 
local minimum in the large scale density distribution. On the assumption that 
the surfaces of such ``spherical'' voids are to be identified with the 
large scale structures hosting the haloes one may attempt to find signatures of 
spin and shape alignments. Voids are not generally spherical, nor is it 
always clear which of several voids a halo is related to. Moreover, using voids as 
defining the environment of a halo does not in itself allow for the important distinction 
between walls and filaments. Alignment detections should therefore be seen as residuals of 
genuine physical alignments with walls or filaments. 
 
This letter poses a number questions for future investigation. The mass segregation
of alignments in filaments is yet to be understood. Detailed merging trees may
give clues to the possible non-linear processes responsible for
this effect.  It will of course be interesting to use MMF to define environments in
catalogues of galaxies such as 2dF and SSDS: this should provide
an important refinement of earlier results of \cite{Trujillo06} and since it 
will unambiguously localise galaxies as either being in walls or in filaments.
%--------------------------------------------------------------------------
%        acknowledgments
%--------------------------------------------------------------------------
\vskip -0.5truecm
\acknowledgments
We thank Pablo Araya for the N-body simulations used in this work and gratefully acknowledge 
discussions with Ignacio Trujillo during the ``Bernard60''conference in Valencia.

\end{document}